# From the microscope to High Performance Computing centers, a national effort toward automated data workflows for microscopy facility users in France

## Authors

Guillaume Gay [1,2] *, Theo Barnouin [1,3], Marc Mongy [1,3], Guillaume Maucort [1,4], Perrine Paul Gilloteaux [1,3] *, Emmanuel Faure [1,2] *

[1] France BioImaging, FBI-Core, Université de Montpellier, CNRS, UAR2057, Paris, France.

[2] Laboratoire d'informatique, de robotique et de microélectronique de Montpellier, LIRMM, Université de Montpellier, CNRS, Montpellier, France.

[3] Nantes Université, CHU Nantes, CNRS, Inserm, BioCore, US16, SFR Bonamy, F 44000 Nantes, France

[4] Bordeaux Imaging Center, Université de Bordeaux, UMS 3420 - CNRS, US4 - INSERM, INRAE, Bordeaux, France

* corresponding authors : guillaume.gay@france-bioimaging.org, perrine.paul-gilloteaux@france-bioimaging.org, emmanuel.faure@france-bioimaging.org

## Abstract

Modern biological microscopy routinely generates large and complex image datasets, including multidimensional, multimodal, and time-resolved acquisitions. While imaging technologies have rapidly evolved, data management infrastructures within microscopy facilities often remain fragmented, relying on heterogeneous local solutions that are difficult to maintain, scale, and integrate with High-Performance Computing (HPC) centers and public data repositories. To address these issues, France BioImaging (FBI), the French national infrastructure for biological imaging, has developed FBI.DATA and the associated BioImage Cloud platform. This initiative aims to provide a coordinated national infrastructure connecting microscopy facilities, centralized storage resources, HPC environments, and public bioimaging archives through interoperable and scalable workflows.

The proposed architecture combines open-source technologies including OMERO for image management, iRODS for distributed data orchestration, Authentik for federated authentication, and emerging standards such as OME-Zarr and REMBI metadata recommendations. The infrastructure is designed to support the complete imaging data lifecycle, from acquisition and transfer to visualization, analysis, sharing, and long-term archiving.

Beyond the technical implementation, this work presents the organizational and governance strategies required to deploy a shared national infrastructure across distributed imaging facilities. We discuss the challenges associated with interoperability, metadata standardization, sustainability, and user adoption, as well as the perspectives opened by tighter integration between imaging data and large-scale computing resources for future AI-driven bioimage analysis workflows.

## Lay Abstract

Modern microscopes can generate extremely large and complex image datasets, especially when recording three-dimensional tissues, long-term live imaging experiments, or multimodal acquisitions. While imaging technologies have advanced rapidly, many microscopy facilities still rely on fragmented local solutions to store, transfer, and analyze these data. This creates major challenges for researchers, including difficulties in sharing data, accessing computing resources, preserving datasets over time, and complying with open-science requirements.

France BioImaging (FBI), the French national infrastructure for biological imaging, has developed the FBI.DATA initiative and the associated BioImage Cloud framework to address these challenges. The goal is to connect microscopy facilities across France to shared computing and storage resources.

The infrastructure combines several open-source technologies to support the complete lifecycle of imaging data, from acquisition at the microscope to storage, visualization, analysis, sharing, and long-term archiving. It also aims to simplify access to large-scale computing resources and facilitate compliance with FAIR data principles, which promote data that are Findable, Accessible, Interoperable, and Reusable.

Beyond the technical aspects, this work also describes the organizational and collaborative strategies required to deploy a shared national infrastructure across multiple institutions. The initiative contributes to ongoing international efforts to improve the management, accessibility, and reuse of microscopy data for the scientific community.

## Introduction

Over the last decade, biological microscopy has undergone a profound transformation. Advances in imaging technologies have enabled routine acquisition of large-scale, high-resolution datasets, including multi-channel 3D volumes, long-term time-lapse recordings (3D+t), and increasingly complex multimodal experiments. While these developments have dramatically expanded the scope of biological questions that can be addressed, they have also exposed a critical weakness in microscopy workflows: data management and exploitation have not kept pace with data production.

As datasets grow in size and complexity, microscopy facilities and end-users face mounting challenges related to storage, computation, metadata curation, and long-term preservation. At the same time, research institutions and funding agencies now impose stringent requirements regarding data stewardship, open science, and compliance with the FAIR principles[1] (Findable, Accessible, Interoperable, Reusable). These constraints place significant pressure on imaging facilities, which often rely on heterogeneous, ad hoc solutions—such as isolated OMERO[2] deployments and local "under-the-bench" servers—that are costly to maintain, difficult to scale, and poorly suited for integration with High-Performance Computing (HPC) resources and public data archives.

This fragmentation has several consequences. It limits interoperability between facilities, hampers user training and support, and ultimately reduces the accessibility and reusability of imaging data associated with scientific publications. From the user's perspective, expectations are shifting toward integrated workflows that minimize technical overhead and provide direct access to computation and data sharing. From an institutional perspective, there is a growing incentive—both economic and ecological—to move away from distributed local storage toward centralized academic computing centers, where data can be stored closer to computational resources and managed more sustainably.

France BioImaging (FBI) is a national distributed research infrastructure that federates 30 biological microscopy facilities across the French territory and hosts about 8000 users every year. To address the structural limitations of local data management and to respond to the increasing demands of modern bio-imaging, FBI has initiated the FBI.DATA project to explore coordinated approaches for integrating microscopy data management workflows at the national scale.

In this article, we present the design principles, technical architecture, governance model, and deployment strategy of FBI.DATA. We describe how this initiative bridges microscopy facilities and national computing centers, enabling FAIR-compliant data management at scale while simplifying the user experience. Beyond its technical implementation, FBI.DATA provides a case study of how national infrastructures may support data-intensive microscopy through coordinated and interoperable data-management strategies. Rather than providing a quantitative benchmarking study, this article focuses on the architectural, organizational, and deployment principles underlying the FBI.DATA initiative.

## Materials & Methods

### The FBI.DATA Initiative

The creation of FBI.DATA team within the national infrastructure FBI was driven by a necessity to restructure how biological imaging data is handled across the national infrastructure. This initiative was founded on a set of strategic objectives designed to transition from fragmented local solutions to a cohesive national strategy.

The primary goal of FBI.DATA is to mutualize IT skills on a national scale. By centralizing expertise and maintenance efforts, the initiative aims to relieve individual imaging facilities of the heavy technical burden associated with complex data management. This centralized organization aims to facilitate maintenance, scalability, and long-term interoperability across participating facilities.

The operational philosophy of FBI.DATA rests on several fundamental principles aimed at sustainability and efficiency:

- **Continuous Evolutionary Maintenance:** The infrastructure commits to ongoing upgrades, version management, and security patches to prevent obsolescence.
- **Coordinated Testing:** New features and updates undergo rigorous alpha and beta testing phases, coordinated across the network to ensure reliability before full deployment.
- **Reinforced Interoperability:** The system prioritizes compatibility with key ecosystem standards and tools, including Zarr[3], iRODS[4], OMERO[2], and RO-Create[5].
- **Decoupling Storage from Application:** A critical architectural decision was to decouple the storage layer from business software applications using iRODS. This ensures flexibility and prevents vendor lock-in.
- **National Data Center Deployment:** FBI.DATA mandates the removal of data storage from local workstations (no disks "under the tables") in favor of deployment in a national data center. This shift may help optimize resource utilization and reduce redundant local infrastructure.

FBI.DATA does not operate in a vacuum; it is explicitly aligned with international efforts and standards. The initiative coordinates with broader European and global ecosystems, such as EuroBioImaging[6] and the BioImage Archive[7], and aligns with concepts like BioImageTown[8]. Furthermore, the data models employed are consistent with established frameworks like ISA[9] (Investigation, Study, Assay) and REMBI[10] (Recommended Metadata for Biological Images), ensuring that data managed within FBI.DATA is standardized and ready for global exchange.

**The Data Life Cycle**

Managing bio-imaging data requires more than storage capacity; it demands a coherent workflow that connects acquisition, analysis, and long-term preservation. The FBI.DATA infrastructure, called BioImage Cloud, was therefore designed to support the complete lifecycle of research data through a structured and continuous pipeline that integrates local acquisition environments with national computing resources and public archives. BioImage Cloud acts as a transport solution designed to facilitate the transfer of imaging datasets from acquisition to a shared cloud storage in HPC centers.

The envisioned lifecycle begins with user onboarding and account creation, ensuring that researchers are properly registered and authenticated within a unified national system. Once access is established, data can be imported from imaging instruments into the infrastructure. This critical step bridges the gap between acquisition at the microscope and centralized data management, relying on local buffer servers to facilitate secure and efficient data transfer while minimizing disruption to acquisition workflows.

After ingestion, the focus shifts to data enrichment and exploitation. BioImage Cloud provides tools for annotation and metadata management, which are essential to ensure that datasets are searchable, interpretable, and reusable. These structured data then enter an active phase of visualization and analysis, where researchers can process large imaging datasets using High-Performance Computing (HPC) resources available through the BioImage Cloud, while limiting the need for large-scale local data duplication.

The final stage of the lifecycle addresses data sharing and dissemination. The infrastructure enables controlled data sharing within research teams and across imaging facilities, supporting collaborative work throughout the project. Upon publication, validated datasets can be transferred seamlessly to public repositories such as the BioImage Archive, ensuring compliance with open science policies and long-term data preservation.

This structured workflow ensures continuity from data acquisition at the microscope to large-scale analysis and long-term archiving in public repositories, via local buffer servers and national computing resources on the BioImage Cloud (Figure 1).

**National Scale Deployment**

The operational rollout of BioImage Cloud follows a centralized yet extensible deployment model. This approach allows the platform and associated hardware infrastructure to scale according to demand while ensuring the harmonization of practices across all participating sites. By centralizing the core architecture, the initiative helps reduce heterogeneity between independently maintained local deployments.

A key component of this deployment is the unified management of access. The system handles the creation and management of accounts, structured hierarchically around Principal Investigators (PIs) and their research teams. Technical implementation relies on robust authentication solutions like Authentik[11] and federation via eduGAIN[12], streamlining access for academic users across institutions.

To ensure the relevance and sustainability of the infrastructure, FBI.DATA operates under a structured governance framework. Strategic direction is overseen by the FBI.DATA Committee , which employs a collegial decision-making process. This governance is not insular; it actively relies on national consultations, such as those with the *Réseau Technologique de Microscopie de Fluorescence*[13] (RTMFM) , and integrates insights from international exchanges notably from Global Bio Imaging[14] or the FoundingGIDE project[15] to ensure the platform remains aligned with global standards.

**Organization of the FBI.DATA Team**

The implementation and sustainability of the FBI.DATA initiative rely on a multidisciplinary team designed to bridge the gap between advanced IT operations and biological research needs. The inter-institutional organization and governance structure of the FBI.DATA initiative are summarized in Figure 3.

The team currently consists of 6 members representing 3.8 full-time equivalents (FTEs) and brings together a diverse range of technical expertise essential for modern data management. It comprises IT engineers, bio-image analysts, data engineers, and metadata specialists. This mix ensures that technical robustness is balanced with domain-specific knowledge relevant to microscopists and biologists. It is reinforced by the contribution of engineers nominated from one of the facilities of the 10 regional nodes, who serve as "DATA referents" to ensure that the local needs are taken into account.

The team functions through a coordinated support model that operates at the national level. A key aspect of this operation is inter-facility coordination, ensuring that solutions and practices are harmonized across the various nodes of France BioImaging. The team maintains continuous interaction with the user base through structured mechanisms such as regular consultations, support ticketing systems, and technical meetings.

Beyond day-to-day operations, the FBI.DATA team actively engages with the broader community to foster adoption and collect feedback. This includes organizing "Data-ckathons" to solve specific data challenges and leading workshops at key scientific events like MiFoBio[16] and in collaboration with the national German BioImaging initiative at TIM[17]. Furthermore, the team relies on beta-user groups to test new features, ensuring that developments meet the practical needs of the field before full deployment.

**Technical Architecture**

The technical foundation of FBI.DATA is built to ensure robustness, scalability, and ease of maintenance. The architecture is designed to support the complete data lifecycle described previously, leveraging a combination of established open-source technologies and modern infrastructure management practices. A detailed overview of the technical architecture and interactions between infrastructure components is provided in Figure 4.

The infrastructure management adopts an "Infrastructure as Code" approach, ensuring reproducibility and streamlined deployment. Physically, the system relies on a National Data Center model—currently comprising two data centers—that hosts data for all distributed nodes. This centralized storage is complemented by local buffer spaces and multiple levels of storage tiering to optimize performance and cost. Figure 2 illustrates this architectural layout, mapping the physical infrastructure to the data flow.

The software stack is composed of complementary technologies chosen for their specific strengths:

- **iRODS[4] (Integrated Rule-Oriented Data System):** This serves as the middleware for distributed data management. Crucially, iRODS facilitates the decoupling of the physical storage layer from the application layer, allowing for flexible hardware evolution without disrupting user services.
- **OMERO[2]:** Used specifically for managing image data

and associated metadata. It provides the user interface through its viewer and offers an API for programmatic access.
- **Omero-Quay[18]:** This component acts as the scheduler/orchestrator, managing jobs and automated tasks within the ecosystem.
- **Authentik[11]:** This solution handles identity and account management, ensuring secure and unified access across the platform.

Several cross-cutting themes guide the technical implementation:

- **Security:** Rigorous protocols are applied to protect sensitive research data.
- **Interoperability:** The system is designed to communicate effectively with other tools and international repositories.
- **Metadata Management:** Emphasis is placed on the enrichment and management of metadata to support FAIR principles.
- **Continuous Upgrade Policy:** The architecture supports a strategy of continuous updates to maintain security and feature parity with the state of the art.

## Results

### User Experience

The infrastructure was designed to simplify user interactions with distributed storage and computing resources. The goal is to minimize administrative friction and technical barriers, while reducing the technical overhead associated with data management.

A major hurdle in distributed infrastructures is often the fragmentation of access. FBI.DATA addresses this by implementing a simplified onboarding process. Through a unified system, accounts are created and managed for all users across the French territory. This centralization allows FBI.DATA to directly handle the management of Principal Investigators (PIs) and their respective teams, ensuring that access rights and group structures are consistently maintained without burdening local administrators.

The user journey is designed as a comprehensive workflow that supports the researcher at every stage:

- **Data Import:** Users benefit from a simplified workflow for importing data from acquisition instruments to the secure infrastructure.
- **Annotation:** Integrated tools facilitate the annotation of datasets, a critical step for structuring information and enabling future retrieval.
- **Visualization and Analysis:** Users can visualize and analyze their images directly within the environment, leveraging the underlying computing power without needing to move large files to local machines.
- **Sharing:** The platform enables flexible sharing mechanisms, allowing seamless collaboration within a research team or across different imaging facilities.
- **Publication and Archiving:** Finally, the journey concludes with streamlined processes for publishing data and archiving it in long-term repositories, ensuring compliance with open science mandates.

### Added Value

The transition to a mutualized national infrastructure brings tangible benefits across the entire bio-imaging ecosystem. By addressing the specific pain points of different stakeholders, FBI.DATA creates a value chain that enhances both scientific productivity and institutional efficiency.

#### For the Research Community

For End-Users, the primary value lies in the abstraction of technical complexity. They benefit from a platform that provides simplicity, stability, and security, allowing them to access FAIR-compliant data that is ready for publication.

Principal Investigators (PIs) gain a potential benefit in data governance. The infrastructure ensures structured management and facilitated sharing, which is essential for collaborative projects. Crucially, it mitigates the risk of data loss associated with staff turnover (e.g., interns or PhD students leaving the lab), ensuring the long-term preservation of research assets while meeting the FAIR requirements of funding agencies.

For PhD Students and Trainees, the platform offers a structured computational environment. It allows them to work within a reproducible framework and facilitates the scaling of their analyses, moving beyond the limitations of local workstations.

#### For Technical and Support Staff

Local IT Staff are relieved of the heavy burden of maintaining complex, isolated storage and software solutions. Instead, they benefit from a national support structure and a technical community that shares best practices. BioImage Cloud has gone through a strict security homologation, which now also brings a label to reinsure the local IT to trust the mutualised management of resources.

Bio-Image Analysts gain unified access to data, enabling the creation of reproducible analysis pipelines that are independent of the data's physical location. Similarly, Data Stewards and Librarians benefit from the standardization of document management, which directly results in higher quality metadata and better indexation.

#### For the Infrastructure and Strategy

At the institutional level, the National Infrastructure achieves significant cost mutualization and greater strategic coherence by avoiding redundant investments in local hardware. Finally, for FBI.DATA itself,the initiative also contributes to international interoperability efforts in bioimaging data management.

## Discussion

### Roadmap and Training Strategy

The long-term success of the FBI.DATA initiative depends as much on community adoption as on technical implementation. At this stage, the initiative should be viewed as an evolving operational framework whose primary objective is to establish interoperable and sustainable national workflows for bioimaging data management. Consequently, the roadmap combines a training strategy with the continuous evolution underlying infrastructure.

To handle the scale of a national infrastructure, FBI.DATA employs a "train-the-trainer" approach. The primary focus is on training designated "DATA referents"—typically IT staff embedded within local imaging facilities. Once trained, these referents act as local experts, responsible for training end-users at their respective sites.

This hierarchical transmission ensures the effective dissemination of best practices throughout the community. Key educational pillars include training on Data Management Plans (DMPs) and ensuring researchers understand FAIR standards and the specific data requirements imposed by funding agencies.

Parallel to the human strategy, the technical roadmap focuses on the continuous evolution of the platform. Future developments aim to enhance performance and integrate next-generation standards. This includes native support for new file formats such as Zarr, the implementation of advanced viewers, and general performance optimizations to handle increasingly large datasets.

## Major Challenges

Implementing a national infrastructure of this magnitude is not without significant hurdles. FBI.DATA faces a convergence of technical, organizational, and economic challenges that must be addressed to ensure long-term viability.

### Technical Heterogeneity and Data Complexity

The primary technical challenge stems from the sheer diversity of data, which varies wildly in terms of file formats and sizes, ranging from simple 2D images to massive multi-dimensional datasets. This diversity is compounded by the complexity of metadata. To be truly useful and FAIR-compliant, metadata must be sufficiently rich, flexible, and accessible, requiring sophisticated management strategies.

Furthermore, the infrastructure must bridge a heterogeneous landscape. It needs to interconnect diverse imaging facilities, stakeholders, and network capabilities , while orchestrating data flows across various types of data centers, including Cloud, High-Performance Computing (HPC), and local storage facilities.

### Sustainability and Human Resources

Long-term maintenance poses a critical challenge that is as much economic as it is technical. Ensuring the durability of the service requires the definition of sustainable economic models, such as determining appropriate billing strategies (e.g., offering the first Terabyte for free).

However, the most pressing sustainability issue is human resources. The complexity of the infrastructure requires highly specialized skills. Securing these skills through permanent staff positions —rather than relying on precarious short-term contracts—is essential to retain expertise and ensure continuous maintenance. It has to be underlined that the mutualisation allows to considerably reduce the investment in human resources compared to a local management (3.8 FTE for 30 facilities and over 8k users).

### Adoption and Support

Finally, the human factor remains decisive. A major hurdle is fostering user adoption by effectively demonstrating the value added by depositing data into systems like OMERO, rather than viewing it as an administrative constraint. This goes hand-in-hand with the operational challenge of scaling user support to a national level, ensuring that thousands of users receive timely assistance.

## Perspectives

As the landscape of bio-imaging continues to evolve, FBI.DATA is actively preparing for the next generation of data challenges. The roadmap focuses on technological innovation and deeper integration of analysis workflows.

A key area of development is the evolution of the storage backend. We are exploring advanced iRODS capabilities to move towards "intelligent datawaves" or concepts similar to "One Data," which would further abstract data location for users. Simultaneously, we are working on expanding OMERO's capabilities to support multi-viewers, moving beyond standard visualization to support more complex, multi-dimensional display needs.

To address the "Big Data" challenge, the adoption of Zarr is being prioritized as a pivotal format. Its ability to handle massive, chunked arrays makes it an ideal standard for the next generation of cloud-ready bio-imaging data.

The future of FBI.DATA lies in reducing the distance between data and computation. Our goal is to achieve a tighter integration where organized data resides as close as possible to computing resources. This proximity is a prerequisite for the efficient integration of Deep Learning workflows. By automating data flows and implementing assisted annotation —leveraging public resources like the BioImage Archive — we aim to potentially reduce manual annotation efforts, supporting future large-scale AI-assisted bioimage analysis workflows.

## Conclusion

The FBI.DATA initiative represents an ongoing effort toward coordinated national bioimaging data management. It represents the transition from a fragmented ecosystem of isolated local solutions to a coherent, sustainable, and FAIR-compliant national infrastructure.

By mutualizing resources and expertise, France BioImaging has developed a shared infrastructure framework that supports the entire research lifecycle, from acquisition to publication. However, the long-term success of this endeavor relies on two critical factors: the widespread adoption by the scientific community and, crucially, continuous institutional support. Sustained investment is vital to maintain the human and technical infrastructure required to adapt to evolving technologies and data volumes.

Finally, FBI.DATA does not exist in isolation. It mirrors and complements similar national initiatives emerging globally, such as those in Germany and across Europe. By aligning its technical choices and governance with these international counterparts, France BioImaging ensures that its national infrastructure acts as a compatible node in a broader, federated global bio-imaging ecosystem, contributing to broader interoperability and open-science initiatives in bioimaging.

### AI-Assisted Writing Statement

Large language model–based tools were used exclusively for language editing, text reformulation, and stylistic improvements during manuscript preparation. No AI-generated content was used for scientific interpretation, data analysis, or the generation of results and conclusions. All scientific content, technical descriptions, and final editorial decisions were reviewed and validated by the authors.

**Acknowledgements**

We sincerely thank all DATA referents across the France BioImaging national infrastructure for their time, expertise, and continued commitment to supporting microscopy data-management activities. Their contributions, feedback, and interactions with end-users have been essential to the development and deployment of the FBI.DATA initiative.

We gratefully acknowledge the institutional and on-premise technical support provided by Olivier Miquel, Jean François Guillaume, Joël Marchand, Julio Mateos-Lingerak, Virginie Georget and the MRI facility, René Marc Mège, Edouard Bertrand, Jean Salamero, Nadine Jaquet, Marine Beraud, and Sylvie Djian.

We also thank Pierre Vincens, Christophe Millien, Sébastien Kita, Stéphane Le Crom, and Michel Chabannes for valuable discussions and advice regarding security and system architecture.

We acknowledge Suzanne Kunis, Katy Wolstencroft, Josh Moore, Matt Hartley, Tom Boissonnet, Aastha Mathur, and the Global BioImaging community for fostering discussions around international community tools, interoperability, and data standards.

We further thank the France BioImaging nodes, the RTMFM and GDR Imabio communities, the IFB MADBOT team, and the “Club des INBS” for their contributions to metadata harmonization, training activities, outreach, and community coordination.

We acknowledge the France BioImaging infrastructure (https://ror.org/01y7vt929), supported by the French National Research Agency (ANR-24-INBS-0005 FBI BIOGEN). The FBI.DATA project benefited from the scientific, technical, and organizational framework provided by France BioImaging, as well as from the engagement of its national community.

This work benefited from State aid managed by the National Research Agency under France 2030 relating to structuring equipment for research / EQUIPEX+ bearing the reference ANR-21-ESRE-0048.

This research used resources of the GLiCID Computing Facility (Ligerien Group for Intensive Distributed Computing, www.glicid.fr, Pays de la Loire, France) and from the ISDM-MESO (Institut de Science des Données de Montpellier,https://isdm.umontpellier.fr/, Occitanie, France ).

Guillaume Gay is a staff engineer at the University of Montpellier. Theo Barnouin is a CNRS engineer. Marc Mongy is an INSERM staff engineer. Guillaume Maucort is a staff engineer at the University of Bordeaux. Perrine Paul Gilloteaux is a CNRS staff engineer. Emmanuel Faure is a CNRS research scientist.

# Figures

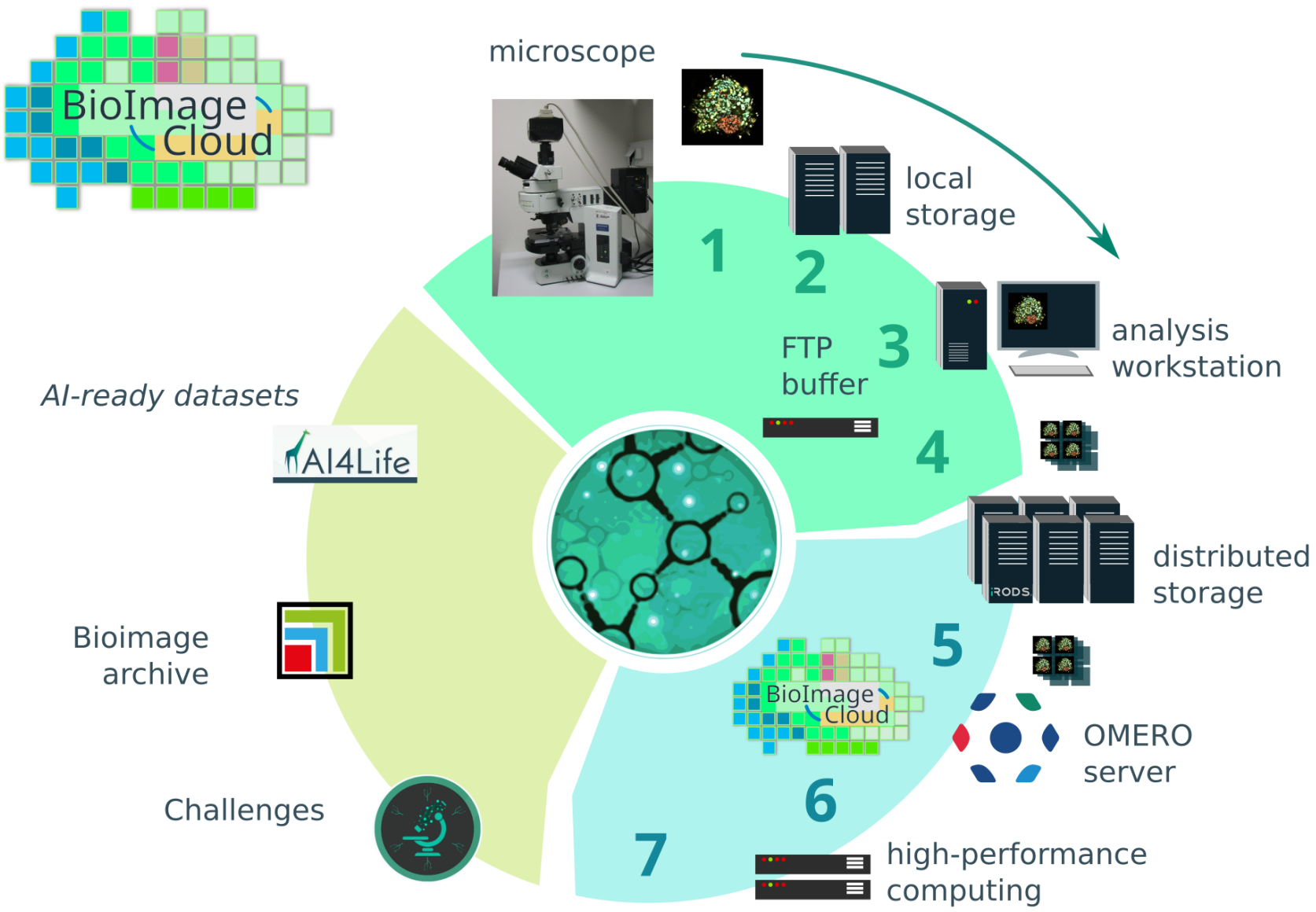

**Figure 1: BioImage Cloud and the imaging data life cycle.** After acquisition and initial curation of microscopy images (1–2), datasets are uploaded to a local transfer server (3). Users can then annotate and import datasets into the BioImage Cloud infrastructure (4). The data are subsequently transferred to appropriate long-term storage resources (5) and made accessible to computing environments for visualization and analysis (6). Finally, datasets may be shared, published in public repositories such as the BioImage Archive, or reused for community challenges and AI training workflows (7).

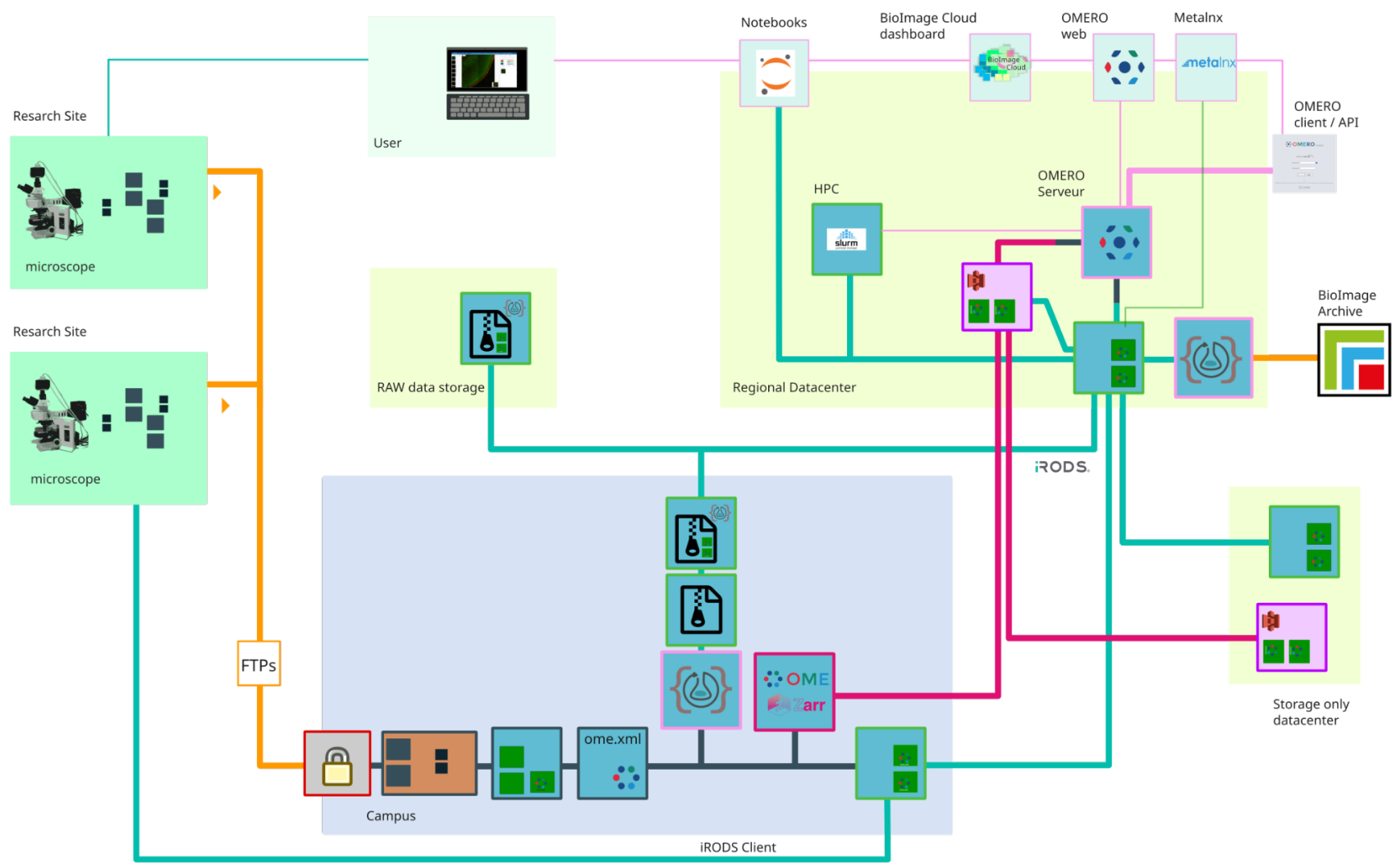

**Figure 2 : Data transport routes within the BioImage Cloud infrastructure.** The modular and asynchronous architecture enables multiple transfer routes for microscopy datasets. Data originating from different research sites (left) can transit through a campus collection service (bottom, blue box), where metadata extraction, archival conversion, and OME-Zarr transformation may be performed. Storage is distributed across multiple data centers, while regional services provide access to datasets through APIs and web-based client interfaces (light green box).

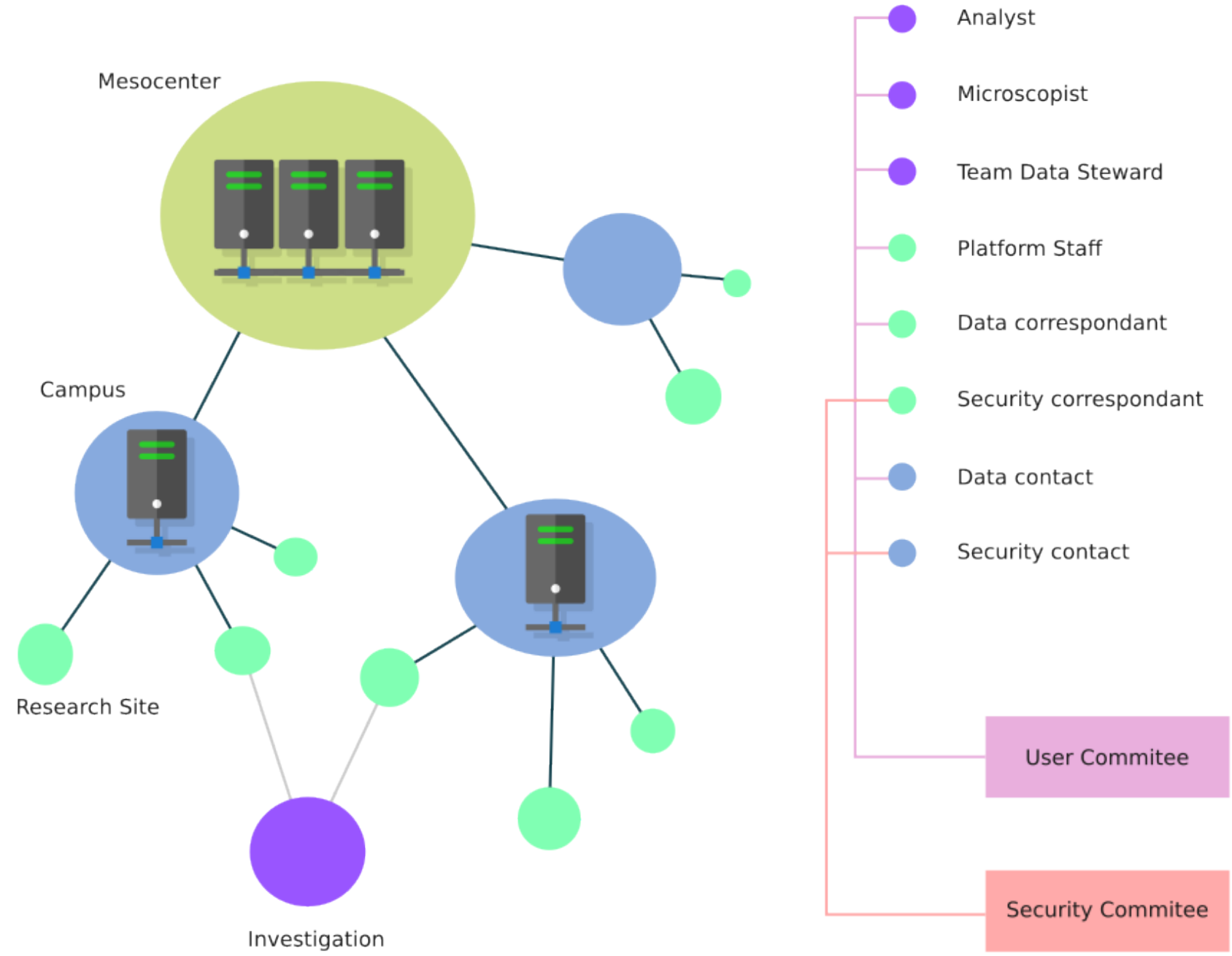

□**Figure 3 : Inter-institutional organizational structure of the FBI.DATA initiative.** The diagram illustrates the interactions between France BioImaging national coordination, regional imaging facilities, local DATA referents, national computing centers, and end-users. It highlights the distributed governance model and coordination mechanisms supporting the deployment and maintenance of shared microscopy data-management workflows at the national scale. Color code: purple, research groups and investigations; green, research sites and microscope locations; blue, campus services and collection servers (FTP buffer systems); cyan, regional or national computing centers (mesocenters). The right panel summarizes the roles associated with each organizational level.

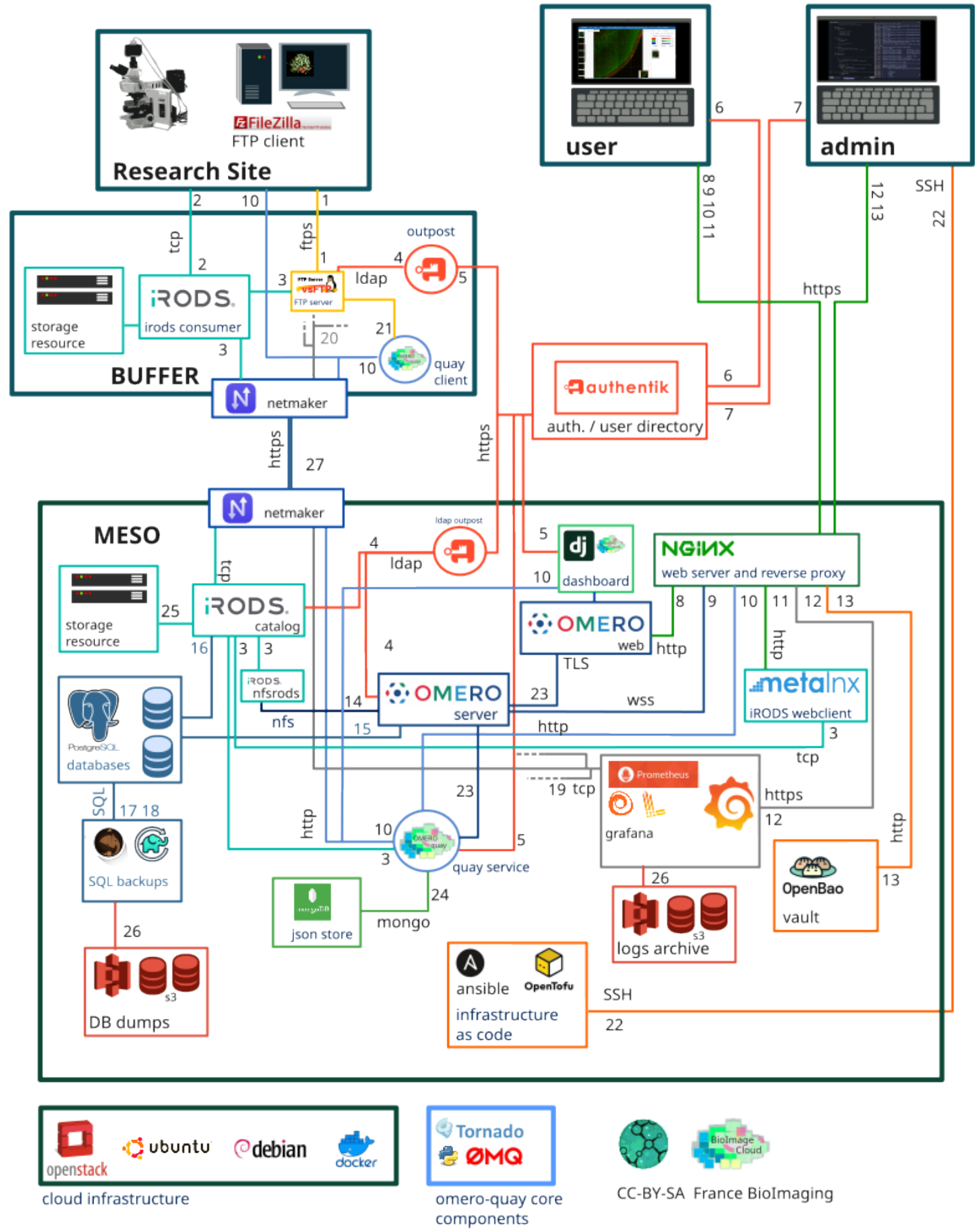

**Figure 4: Detailed technical architecture of the BioImage Cloud infrastructure.** The diagram presents the main software and infrastructure components involved in microscopy data-management workflows, including local acquisition systems, buffer storage, regional data centers, distributed storage orchestration through iRODS, image

management with OMERO, authentication services, and integration with public repositories. The figure also illustrates the interactions between these components within the FBI.DATA deployment framework, as well as ancillary services such as logging systems, databases, secure vaults, and network communications between infrastructure components.